\documentclass[a4paper,10pt]{article}

\usepackage[utf8x]{inputenc}
\usepackage[T1]{fontenc}
\usepackage[english]{babel}
\usepackage{amssymb}
\usepackage{amsmath}
\usepackage{amsthm}
\usepackage{fullpage}
\usepackage{verbatim}

\newtheorem*{theo}{Theorem}
\newtheorem*{question}{Question}

\newcommand{\Ra}{\Rightarrow}

\newcommand*\lon{%
       \mskip1mu
        \relax
        {:}%
        \mskip1mu
        \relax
}

\title{Conditional and unconditional information inequalities:
\\ an algebraic example
\footnote{Authors are supported by ANR NAFIT grant 008-01}}

\date{\today}

\author{
Tarik Kaced\\{LIRMM, Univ. Montpellier II}
\and  Andrei Romashchenko\\{LIRMM, CNRS \& Univ. Montpellier II}
}

\begin{document}
\maketitle

\begin{abstract}

We provide a simple example showing that some conditional information
inequalities (even in a weak form) cannot be derived from unconditional
inequalities. 

\end{abstract}

\section{Introduction}

Three conditional information inequalities were considered
in~\cite{ZY97,Matus99,condineq}:
$$
\begin{array}{rlclcl}
 (1)&\label{bla}I(A\lon B|C) = I(A\lon B)=0 &\Ra&
    I(C\lon D) \le I(C\lon D|A) + I(C\lon D|B) \\
 (2)&I(A\lon B|C) = I(B\lon D|C)=0 &\Ra&
    I(C\lon D) \le I(C\lon D|A) +I(C\lon D|B)+I(A\lon B)  \\
 (3)&I(A\lon B|C) = H(C|AB)=0  &\Ra&
    I(C\lon D) \le I(C\lon D|A) + I(C\lon D|B) + I(A\lon B)  \\
\end{array}
$$
Here $A,B,C,D$ are random variables (with a finite range), $H(U|V)$ stands for
conditional Shannon entropy of $U$ given $V$, and $I(U\lon V|W)$ is the amount
of mutual information in $U$ and $V$ when $W$ is known.
\begin{question}
Can these conditional inequalities be derived from their unconditional
versions?
\end{question}
For example, inequality~(1) could be a consequence of an unconditional
inequality (with ``Lagrange multipliers'') 
     $$
I(C\lon D) \le I(C\lon D|A)+I(C\lon D|B) + \kappa[I(A\lon B|C)+I(A\lon B)]
     $$
if the latter were true for some $\kappa>0$. This is not the case, as shown
in~\cite{condineq} (for all three inequalities (1-3)). In this note we consider
a statement that can be considered as a conditional inequality that is weaker
than both~(1) and~(3):\footnote{The condition $I(A\lon B|D)=0$ is not needed to
derive~(1) and~(3). We added it to make the statement weaker (and the result
below stronger).}
$$
I(A\lon B|C) = I(A\lon B|D) = H(C|AB) = I(C\lon D|A) = I(C\lon D|B) = 
I(A\lon B)=0
\Ra I(C\lon D) = 0
$$

We give a simple algebraic example that shows that even this weak statement is
essentially conditional, therefore providing an alternative simple proof of
the conditional nature of inequalities~(1) and~(3).

\begin{theo}
There is no constant $\kappa$ such that
\begin{equation*}
I(C\lon D) \le \kappa\left[I(A\lon B|C) + I(A\lon B|D)+ H(C|AB) + 
I(C\lon D|A) + I(C\lon D|B) + I(A\lon B)\right] \eqno(*)
\end{equation*}
for all random variables $A,B,C,D$.
\end{theo}

\section{Example}

Consider a quadruple $\langle A,B,C,D\rangle_q$ of geometric objects on the
affine plane over the finite field $\mathbb{F}_q$ :

\begin{itemize}

\item First choose a random non-vertical line $C$ defined by the equation
$y=c_0 + c_1x$ (the coefficients $c_0$ and $c_1$ are independent random
elements of the field); 

\item then pick uniformly at random a parabola $D$ in the set of all
non-degenerate parabolas $y=d_0+d_1x+d_2x^2$ (where
$d_0,d_1,d_2\in\mathbb{F}_q, d_2\neq 0$) that intersect $C$ at exactly two
points;

\item and call two intersection points $A$ and $B$ in a random order.

\end{itemize}
One can easily compute both sides of the inequality~$(*)$ for these $A,B,C,D$,
and get the following inequality:
$$
1-\log_2{\left(\frac{q}{q-1}\right)} \le  4\kappa\log_2
{\left(\frac{q}{q-1}\right)} 
$$
This leads to a contradiction: the left-hand side is $1-O(\frac1{q})$ and the
right-hand side is $O(\frac1{q})$ for large $q$.

\paragraph{Informal explanations:}

The mutual information between the line and the parabola is approximately $1$
bit because $C$ and $D$ intersect at two distinct points iff the corresponding
equation discriminant is a non-zero quadratic residue, which happens almost
half of the time. Other information quantities vanish since the involved random
geometrical objects are almost independent. 

It is in fact easy to check that they are all equal to
$\log_2{(\frac{q}{q-1})}$. For instance, given $A$ there are $q$
equiprobable lines $C$. Given $A$ and $D$, there are $q-1$ equiprobable lines
since now the tangent to $D$ at $A$ is excluded ($A\neq B$).  Hence $I(C\lon
D|A) = \log_2{(q)} - \log_2{(q-1)}$. All other computations are similar.

\section{Historical comments and motivation}

To every quadruple of discrete random variables corresponds an \emph{entropy
vector} consisting of the fifteen entropies of all possible non-empty subsets
of variables (in some fixed order). A vector is called \emph{entropic} if it is
the entropy vector of some tuple of random variables.  The set of all entropic
vectors has a very complex structure (e.g., it is not even closed). However,
the closure of the set of all entropic vectors is easier to study: it is a
convex cone. This closure is called the cone of \emph{asymptotically entropic
vectors}.

Some bounds for the cone of asymptotically entropic vectors are simple and
well-known.  For instance, these vectors satisfy \emph{Shannon inequalities},
i.e., all non-negative linear combinations of the basic inequality 
$$H(X,Z) + H(Y,Z) - H(X,Y,Z) - H(Z)  \ge 0.$$
This inequality means that the conditional mutual information $I(X\lon Y|Z)$ is
non-negative.

N.~Pippenger asked in 1986 a general question \cite{Pippenger} : does there
exist any linear inequality for entropies other than Shannon inequalities?  The
first inequality of this kind was conditional \cite{ZY97} and appeared in 1997,
followed by a similar unconditional one \cite{ZY98} a year later. Quite many
(actually, \emph{infinitely} many) other inequalities have been found since
then. In fact,  the set of all valid information inequalities cannot be reduced
to any finite number of ``basic'' inequalities: F.~Mat\'{u}\v{s} proved that
the cone of asymptotically  entropic vectors for quadruple of random variables
is not polyhedral \cite{Matus-inf}.
 
Conditional equalities (1-3) express other subtle properties of entropic
vectors. These inequalities imply conditional independence inference rules
(which can not be deduced from Shannon's inequalities). These rules were
extensively studied by several authors, see a survey by M.~Studený in
\cite{StudenyThesis}. 

In \cite{MMRV} it was asked if any conditional information inequality (in
particular (1)) can be obtained as a direct consequence of some unconditional
inequality. The negative answer was given in \cite{condineq}. In the present
paper we give a simplified proof of this result, in a slightly stronger form
for~(1) and~(3).

\section*{Acknowledgements}
The authors thank Alexander Shen for encouragements to write this note and for
many fruitful comments and discussions


\begin{thebibliography}{10}

\bibitem{Pippenger} N.~Pippenger, 
What are the laws of information theory.
1986 Special Problems on Communication and Computation
Conference, Palo Alto, California, Sept. 3-5 1986.


\bibitem{ZY97} Z.~Zhang and R.~W.~Yeung. 
A non-Shannon-type conditional information inequality. 
IEEE Transactions on Information Theory, 43(1997), pp. 1982--1986.

\bibitem{ZY98} Z.~Zhang and R.~W.~Yeung. 
On characterization of entropy function via information inequalities. 
IEEE Transactions on Information Theory, 44(1998), pp. 1440--1450.



\bibitem{Matus99} F.~Mat\'{u}\v{s}.
Conditional independences among four random variables III: final conclusion.
Combinatorics, Probability \& Computing, 8 (1999), pp.~269--276. 

\bibitem{Matus-inf} F.~Mat\'{u}\v{s}. Infinitely many information inequalities.  
IEEE ISIT 2007, pp.~41--44.

\bibitem{StudenyThesis} M.~Studen\'{y}.
On mathematical description of probabilistic conditional independence 
structures,
thesis for DrSc degree, Institute of Information Theory and Automation,
Academy of Sciences of the Czech Republic, Prague, May 2001, 192 pages.

\bibitem{MMRV} K.~Makarychev, Yu.~Makarychev, A.~Romashchenko, N.~Vereshchagin.
A New Class of non-Shannon Type Inequalities for Entropies.
Communications in Information and Systems. 2 (2002) No. 2, pp.~147--166.

\bibitem{condineq} T.~Kaced, A.~Romashchenko. 
On essentially conditional information inequalities. 
IEEE ISIT 2011, pp.~1935--1939.

\end{thebibliography}
\end{document}